# Fissile Material Detection using Neutron Coincidence Counters

Sebastian Ritter

Ken and Mary Alice Lindquist Department of Nuclear Engineering

The Pennsylvania State University, University Park, PA 16802

*Abstract* — A paper study is conducted on the detection of fissile material using neutron coincidence counters. Three unknown independent properties of assays of fissile material assays are identified to be the effective mass, the (alpha,n) production rate and the neutron multiplication rate. Singlet counting can only be implemented for very well-known assay samples. Doublet counting may be employed with calibration curves which reduce the number of unknowns from three to one. Multiplicity counting theoretically enables assay mass determination without calibration curves. Coincidence counting is most widely applied via active and passive well coincidence counters. Passive neutron well counters are well suited for determination of 240 Pu masses. Active neutron well counters are well suited for determination of fissile material masses. Active well counters are found to have relative mass errors of 4% for a highly enriched uranium (HEU) sample of mass of 1kg and 8% for a mass of 50g. Thermal neutron interrogation sources feature smaller relative errors in small assays than fast neutron interrogation sources.

## I. Introduction

The NPT Article III lays out that each non-nuclear weapons state (NNWS) must have safeguards in place on all its peaceful nuclear activities. Safeguard agreements are to be negotiated with the IAEA. IAEA safeguards are tasked with the timely detection of diversion of nuclear material and to deter of such diversion by "risk of early detection" [1]. INFCIRC/153 Part 1 lays out rights and responsibilities of member states. A state shall "establish and maintain a system of accounting for and control of all nuclear material subject to safeguards", or SSAC for short [1]. INFCIRC/153 Part 2 is concerned with technical aspects of safeguards.

The International Atomic Energy Agency (IAEA) is tasked with verifying a NNWS's accountings and with verifying a state's non-diversion of nuclear material. This is done via independent measurements of SSAC declared assay values [1].

Active and passive neutron coincidence counting are the primarily implemented methods to verify declared values of Plutonium and Uranium assays.

## II. Theory

Plutonium and Uranium allays are measured in different ways depending on their spontaneous fission rates. Plutonium isotopes with even mass numbers are known to exhibit spontaneous fission rates that are at least three orders of magnitude greater than isotopes featuring odd mass numbers. Likewise, Uranium isotopes featuring even mass numbers are known to exhibit spontaneous fission rates that are at least one order of magnitude greater than Uranium isotopes featuring odd mass numbers.

The high spontaneous fission rates in Plutonium samples allow for the measurement of such samples via passive detection systems. Uranium samples are measured with active detection systems utilizing neutron interrogation sources to induce fission.

Other dominating factors in free neutron production in fissile materials is the induced fission process and the (alpha,n) reaction. The induced fission process will introduce a multiplication of coincident neutrons within the sample. The (alpha,n) reaction will result in an increase of total free neutrons in the sample.

Fissile neutrons from different isotopes cannot be distinguished in detection systems [2]. For that reason, it is convenient to define an equivalent mass for Plutonium fuels. The 240 Pu equivalent mass is the mass of 240 Pu that would give an equivalent neutron count rate as the count rate obtained from a mix of all the even numbered Plutonium isotopes. This mix is a linear combination of all the even numbered Plutonium isotopes. The coefficients in this linear combination depend on the spontaneous fission half-lives, neutron multiplicities and detector responses when measuring these multiplicities. The choice of 240 Pu as a reference mass in Plutonium containing samples is made because 240 Pu is a major component in both low and high burn-up reactor grade Plutonium. Ensslin [3] calculated a linear combination for a shift register circuitry in equation (1).

$$^{240}_{eff}Pu = 2.52\ ^{238}Pu + \ ^{240}Pu + 1.68\ ^{242}Pu \quad (1)$$

240 Pu eff mass can be used to calculate the total mass of Plutonium if the isotopic composition of the sample is known. For Uranium samples, 238 U is the single dominant spontaneous fission source. The 238 U eff mass may be calculated similarly to the 240 Pu eff mass. However, due to the low spontaneous fission rate of 238 U, Uranium assays are actively interrogated in most applications.

There are three unknown independent properties of fissile material samples for safeguard applications. The 240 Pu eff mass, the (alpha,n) production rate and the neutron multiplication in the sample [3]. The multiplication factor depends on physical sample characteristics such as density, geometry, and material type. The (alpha,n) production rate is strongly dependent on impurities, in particular low-Z impurities, many of which feature large (alpha,n) cross sections.

The detection of free neutrons is called singlets counting. The detection of two coincident neutrons is called doublets or coincident counting. The detection of three coincident neutrons is known as triplet counting or as multiplicity counting.

Under knowledge of two of the three unknowns, singlets counting allows for the determination of the 240 Pu eff mass. Under knowledge of one of the three unknowns, doublets counting allows for the determination of the 240 Pu eff mass. Calibration curves provide knowledge of two of the three unknowns. For triplet counting, none of the three unknowns are theoretically required to be prior knowledge to determine the 240 Pu eff mass. The big disadvantage of the triplet counting method is that the sample is assumed to follow a so-called point model. Thus, while this technique does not require calibration curves, they are still made to determine correction factors for the point model [3].

If a sample is a pure actinide metal, the (alpha,n) production rate will be zero due to the lack of low-Z chemical compounds such as oxides. If a sample is sufficiency small, it can be non-multiplying. If both of these two simplifying material factors are known to be true, simple passive single neutron count measurements will yield the 240 Pu eff mass via equation (2) [2].

$$S = k * {}^{240}_{eff}Pu \qquad (2)$$

S is the single neutron detection rate and k is a calibration factor which is found by using samples of known mass and isotopic composition.

Likewise, for rare instances of very large Uranium assays, equation (2) may be written for passive single neutron measurements as

$$S' = k' * {}^{238}_{eff}U \qquad (3)$$

For single neutron detection, IAEA safeguards are burdened to gain knowledge of the chemical form and isotopic composition for a quantitative analysis. In non-metallic samples, (alpha,n) reactions render the measurement of fissile materials nearly impossible [2]. If an (alpha,n) neutron source was implemented in a simple neutron counting system for interrogation, the induced fission signal would be masked by the dominating (alpha,n) neutron source.

Counting two time-correlated neutrons allows for the discrimination of (alpha,n) reaction induced neutrons from fission neutrons to a large extent.

III. IMPLEMENTATION OF NEUTRON COINCIDENCE COUNTERS

Neutron coincidence counters are currently implemented and developed for safeguards applications in two ways. First the widely applied neutron coincidence well counters and second liquid scintillator-based counters [4]. These two technologies differentiate foremost in their detector material and signal processing techniques. Because of their relative novelty in safeguards, only neutron coincidence well counters are discussed. Neutron multiplicity counting is another powerful neutron coincidence measurement technique. This method is beyond the scope of this literature review paper.

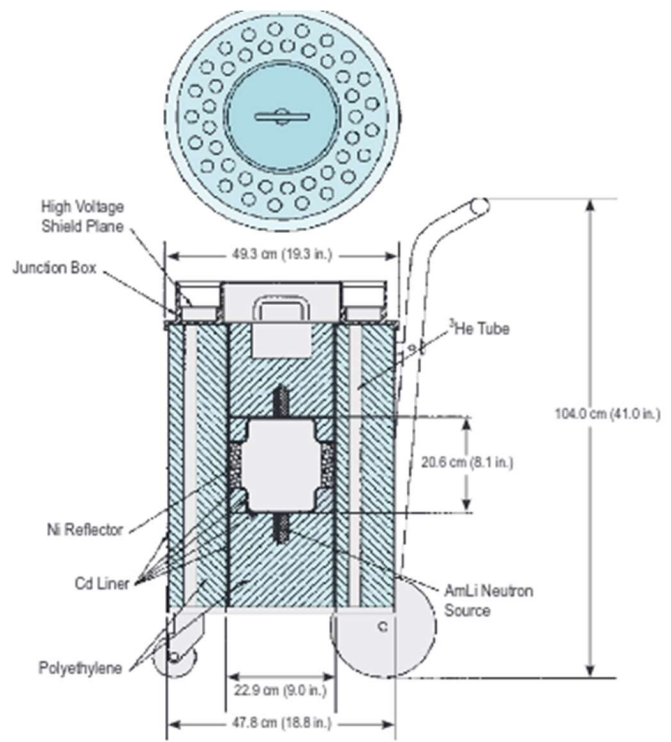

**Fig. 1.** Schematic image and pictorial description of the Canberra JCC-51 active and passive well coincidence counter. The upper image shows a top view of Canberra JCC-51 and the lower image shows a cross section of the well coincidence counter. The sample chamber is seen in the middle of the counter. The transportation crate is also shown. Credit: Canberra [5]

*A. Description of 3 He Well Counters*

3 He well counters consist of many 3 He detector tubes which are situated in the walls of a cylindrical well detector. The dimensions of such a well detector are typically 1 meter in length and half a meter in diameter. The walls are filled with polyethylene slabs to moderate high energy fission neutrons to thermal energies which offer a greater absorption cross section in the low density 3 He gas-filled proportional detector tubes.

A sample of nuclear material important to safeguard analysis is inserted into the well detector sample chamber and measured for a certain period.

Plutonium assays are measured in passive well counters and Uranium assays are measured in active well counters due to the low spontaneous fission rate of Uranium, rendering an active interrogation neutron source a necessity.

The Canberra model JCC-51 is most widely used in safeguards applications. A schematic image of the Canberra JCC-51 is shown in Figure 1. The unit is of cylindrical shape with 42 [5] 3 He detection tubes and two 3 He detection rings. The 3 He tubes are surrounded by polyethylene sheets to increase detection efficiency. The unit is portable and transportable with a crate. Including the crate, the unit is 104cm in height and 49cm in diameter. The sample chamber size is of cylindrical shape and 20.6cm by 22.9 cm in size with polyethylene plates fully inserted. With all polyethylene plates removed, the size is 35.1cm by 22.9 cm [5]. The polyethylene sheets are used to adjust the height of the sample inside the sample chamber. The unit has a mass of 125 kg [5]. When all sheets are inserted, absolute detection efficiency is increased due more neutrons reaching the button ends of the 3 He tubes. At the same time measurement precision is decreased because the neutron flux from the Americium-Lithium neutron source (AmLi) is increased [5].

Cadmium is wrapped around the detector to reduce user exposure and to lower background counts. The unit uses an AmLi source with a neutron emission rate of 5*10^4 neutrons per second [5]. When the AmLi source is removed, the detector may be operated in passive mode. The unit offers a passive and active detection mode which is switched by inserting or removing the AmLi source. In active mode, the unit offers a thermal and a fast mode.

In thermal mode, the internal cadmium wrapping as well as the cadmium plates at the end plugs are removed. In the fast mode, these plates are inserted. The main purpose of these cadmium plates is to prevent neutrons in the polyethylene moderation sheets from diffusing back into the sample chamber and inducing fissions [5] [6]. The consequence is a lower sample sensitivity to self-multiplication and a lower sample sensitivity to sample density and enrichment variations [5]. Conversely, in thermal mode, assay samples will be very sensitive to their self-multiplication. This sensitivity is arising from the large amount of polyethylene thermalized neutrons diffusing back into the sample [5] [6]. The samples are also very sensitive to densities and to enrichments due to the lower sample penetration capability of the thermal neutrons when compared to fast neutrons [5] [6].

The lifetime of a neutron in the detector is described by the neutron die-away time τ. Equation (4) describes the number of neutrons in the detector as a function of time t, the initial number of neutrons $N_0$ and the die-away time [3]. N. Ensslin found a gate time of approximately 1.26 τ to minimize the measurement of accidental coincidences [3].

$$N(t) = N_0 * e^{-\frac{t}{\tau}} \qquad (4)$$

The absolute neutron detection efficiency of a 3 He counter is strongly depended on the neutron energy spectrum incident on the detector. L' Annunziata et al. [2] state that if the detection efficiency is $d$ for single events, it is $d^2$ for doublets and $d^3$ for triplet events.

*B. Statistics of He3 Well Counters*

Shift register electronics are used to obtain the real coincidence rate R and the accidental coincidence rate A. The real coincidence rate R is the rate of detection of coincident neutrons which are originating from true fission events. The accidental coincidence rate A are random coincidental neutron detections. Two simple statistical models may be used to differentiate A from R which are made use of in the shift register electronics [3].

First, is the distribution of time intervals between detected coincidental events. I(t) shall be the probability of detecting a coincidence event which has a time interval of t. For a total coincidence count rate of T the distribution of intervals may be expressed as equation (5) [3].

$$I(t) = T * e^{-t*T} \qquad (5)$$

The probability is expected to follow an exponential decline in interval length. Plotting this exponential decline in a log-lin graph, the exponential decline becomes a straight line. The accidental count rate is not expected to follow this exponential law since it is not more likely to occur immediately after the first event. An experimental result is shown in Figure 2. Fitting of straight lines to the low and high t regions of the Figure 2 allows for discrimination of R and A.

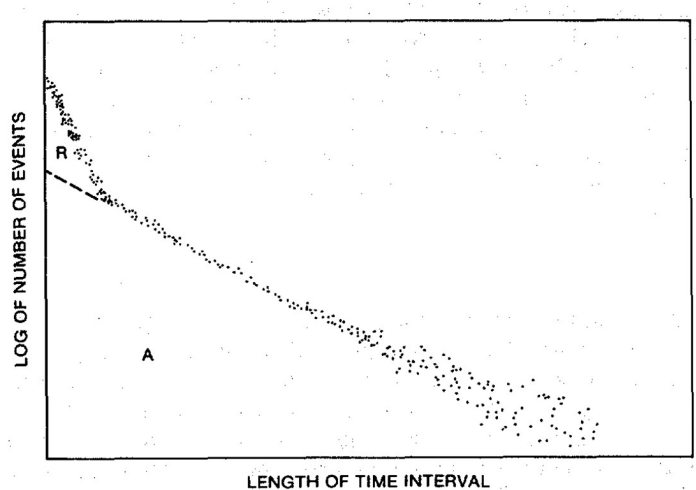

**Fig. 2.** Plotting of an experimental result of the probability of detecting a coincidence event which has a time interval of t associated between them. The plot is a log-lin graph. An exponential decline of the real coincidences, denoted as R, is visible as a straight line. The exponential decline is described in equation 5. Accidental coincidences, denoted as A, feature very long time intervals. Accidental coincidences can be distinguished from real coincidences. Credit: N. Ensslin [3]

Second, is the Rossi-Alpha distribution. This is a distribution in time of detected neutrons following an initiating neutron detection. This distribution is created by staring a clock at time t = 0 with measurement of a neutron detection event. Each arriving neutron detection is time stamped and saved in a bin corresponding to its arrival time relative to the first detection event. The clock is restated once all bins are filled. A typical bin can be 1024 entrees long. The distribution is given by equation (6) [3].

$$S(t) = R * e^{-\frac{t}{\tau}} + A \qquad (6)$$

S(t) is the probability of detection as a function of time following the beginning detection event. $\tau$ is the neutron die-away. R and A are respectively the real and accidental coincidences. The Rossi-Alpha distribution is shown in Figure 3 in a lin-lin plot. A and R can be differentiated by mathematical subtraction.

*C. Electronics of He3 Well Counters*

The value R may be determined with shift register electronics. Two detection events are defined to be coincident when they occur within the gate time, denoted as G. There are at least two basic implementations of counting such coincidences.

First, an incoming neutron detection triggers a gate opening at time t = 0. Each incoming neutron detection is counted while the gate is open. The gate closes at time t = G and the total number of counts within G is read out. This method creates a deadtime of time length G since the gate is busy counting neutron events during the time interval G. [ref essiln]

Second, shift register electronics may be employed to remove the deadtime limitation. Furthermore, such electronics enable to compare every detection event with every other detection event occurring within the gate time G. Shift register electronics consist of a shift register, an up/down counter, a long delay and two scalars, one for R+A counting and one for A counting [3]. Incoming events are stored for time G in the shift register which is a series of flip-flops, driven by a clock that typically runs at 1 MHz or at 1 μs per shift. A 32bit shift register has 32 stages, yielding a gate time of 32 microseconds. An up/down counter is attached to the shift register which counts the number of events in the shift register. This is done by counting up when an event enters the shift register and counting down when an event leaves the shift register. The up/down counter may be read by a scalar to provide for the resulting number of events stored in the shift register. A "R+A scalar" [3] reads the number of random plus accidental events within time interval G following a triggering initial event at t = 0. A many gate lengths G delayed "A scalar" [3] reads the accidental rates. Note that the Rossi-Alpha distribution shows a very low probability of detecting real coincidences at many gate lengths after the triggering event at t = 0. A schematic image of shift register electronics is shown in Figure 4.

An example is given to illustrate shift register electronics. An incoming neutron pulse both travels through the shift register and immediately triggers the "R+A scaler". As it travels through the shift register, four more events follow. Of these four events one is a real coincidence and three are accidental coincidences. The up/down counter registers four events and the "R+A scalar" reads four. The delayed "A scalar" reads the shift register long after the real coincidence left the sift register. New events were recorded in the shift register in the interim of the delay. These events, which occur a long time after the initiating event at t = 0, are likely to be only accidental coincidences. This is because the Rossi-Alpha distribution shows accidental coincidences to have an equal probability of occurring in the "R+A scalar" or in the "A scalar". Thus, the number of events read by the "A scalar" is likely three. Subtracting the "R+A scalar" from the "A scalar" yields R, the number of real coincidences following the initiating event at t = 0. A pre-delay may be inserted before the shift register to filter out pulse-pileup and deadtimes biases.

Once R is determined, 240 Pu eff and 238 U eff mass may be calculated.

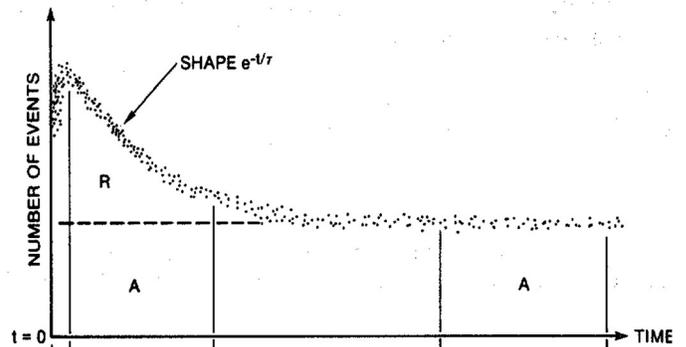

**Fig. 3.** Plotting of the probability of detection S(t) as a function of time from an initiation detection at t = 0. The plot is a lin-lin graph. The function follows a Rossi-Alpha distribution. An exponential decline in real coincidences, denoted as R, is visible. Accidental coincidences, denoted as A, have an equal chance of occurring immediately after t = 0 to any later time. Accidental coincidences can be distinguished from real coincidences. Credit: N. Ensslin [3]

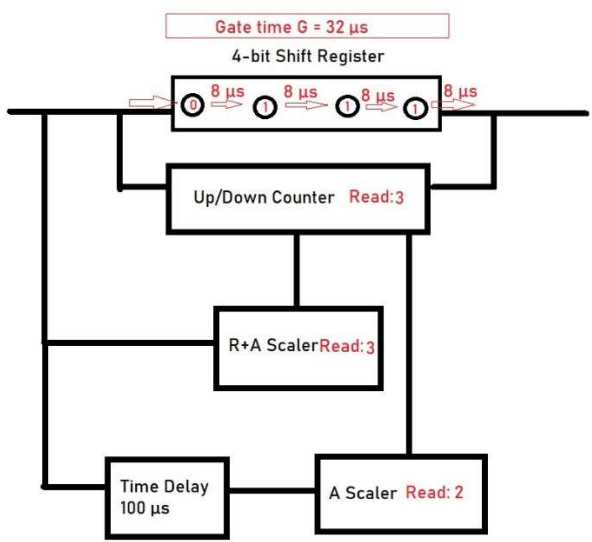

**Fig. 4.** Schematic image of shift register electronics. Shown are a shift register, an up/down counter, a long delay and two scalars. The A scalar is time delayed by 100 μs. A gate length G of 32 μs is shown for a 4-bit shift register yielding 8 μs per flip flop operation. An incoming pulse at t = 0 triggers the "R+A Scaler" and is recorded as 1 in the shift register. Immediately two other events follow, with a non-detection at the last 8 μs step yielding the output 0111 of the sift register. This is read as 3 by the Up/Down counter and as 3 for the "R+A Scaler". Following a (100 – 32) = 68 μs time delay, the "A scaler" is likely to read 2 if two events were accidental coincidences and one event was a real coincidence event. After descriptions of N. Ensslin [3].

*D. Passive Well Counters*

Typically, passive well counting is conducted only for Plutonium samples and with the help of a calibration curve [2]. This curve is generated using Plutonium mass standards and fitting the coincidence rate of two neutron detections with the known Plutonium mass values. The result is the coincidence rate as a function of 240 Pu eff mass. This calibration curve enables to associate the R value with 240 Pu eff mass values.

The calibration curve method reduces the three independent unknown properties of fissile material samples to just one. With measurement of R, the 240 Pu eff mass can be determined. However, due to the reduction of two dependencies, the calibration curve strongly depends on the sample (alpha,n) production rate and on the multiplication value [2]. This has the consequence that the utilized calibration standards are required to be representative of the assay. Some of the dependencies include and are not limited to the material geometry, chemistry, moisture, impurities, phase state and density [2]. Assay samples best suited for the calibration curve method are those that are intended to be reproduced, such as assays from reprocessing plants [2].

If the (alpha,n) production rate of a sample is known, the so-called alpha method can be used [2]. In this method, both the detected single and double neutron detection values are used. Using these values, the multiplication factor is determined eliminating one of the three unknowns. A new calibration curve is then obtained by fitting the coincidence rate of double neutron detections with known Plutonium mass values. The advantage of this technique is that the new calibration curve is now independent of a sample's multiplication factor [2]. If the (alpha,n) production rate of the sample is not well known or poorly reproducible between standards and assay samples, this method is very limited in its applications. Metals and oxides reactor fuels are often measured with this method allowing for tolerance against fuel swelling and other radiation induced material changes [2] [3].

*E. Active Well Counters*

Some passive well counters are convertible into active well counters [5]. Two neutron sources are attached on the top and on the bottom of the well counter [5].

Apart from special cases of large assay samples, Uranium is exclusively measured with active detection systems due to their low spontaneous fission rates [2] [3]. Typically, an AmLi source is used for three reasons [2] [3] [5]. First this source has a low gamma to neutron production ratio. Second, it features a random neutron generation originating from radioactive decay. Third, its neutron energy spectrum has the majority of the neutrons at energies below the fission neutron threshold of 238 U fission.

The measurement method of an active well counter is comparable to that of a passive well counter [2] [3]. In the calibration curve method, double coincident neutron detection rates are measured for interrogated standard Uranium samples. Following calibration, the double coincident neutron detection rate is then related to the 235 U mass [6].

## IV. PERFORMANCE OF AN ACTIVE WELL COUNTER IN HEU APPLICATIONS

F. Ferrari et al. [6] provide a paper on active neutron interrogation methods of well coincidence counters for highly enriched uranium (HEU) samples. The goal of this paper is to conduct a calibration for an active well counter using 235 U standards. Following calibration, the researchers attempt to identify a compromise value for the assay sample measurement time which yields an HEU mass value within the declared sample mass value and within an acceptable statistical significance.

The researchers utilized the widely utilized CANBERRA models JCC-51 and JSR-12 for sample detection and sample analysis. Two modes of neutron source operation are noted, the thermal neutron mode and the thermal neutron mode. The thermal neutron mode offers increased induced fission cross sections in HEU with the drawback of low sample penetration and low accuracy due to an increased multiplicity dependency of generated neutrons. The fast neutron mode offers higher accuracy, higher sample penetration, but lower fission cross sections leading to lower neutron detection count rates. HEU and homogeneous samples benefit from the fast mode.

The standard calibration samples all consist of HEU in the form of Uranium Dioxide ($UO_2$). Samples were of cylindrical shape with the same radius but varying lengths.

The calibration method is used for quantitative analysis. Calibration is conducted on HEU samples of four different enrichments and each enrichment is measured for 5 to 7 different mass samples. Sample count time is measured with cycles. Each cycle is set at a time of 100s. Sample counting is aborted when an arbitrarily chosen target precision of 0.5% for the thermal mode and of 1% for the fast mode is reached. The results are eight calibration curves: 4 for each enrichment with two curves each for the thermal and fast mode. Data points are fitted to the calibration curve given in equation (7) [6].

$$R = a * m_{235\,U}^{b} \qquad (7)$$

R is the real coincidence detection rate, $m_{235\,U}$ is the mass of the standard calibration fissile sample. The coefficients a and b, along with their uncertainties, are determined. The calibration curves are shown in Figure 5 for the thermal mode and in Figure 6 for the fast mode.

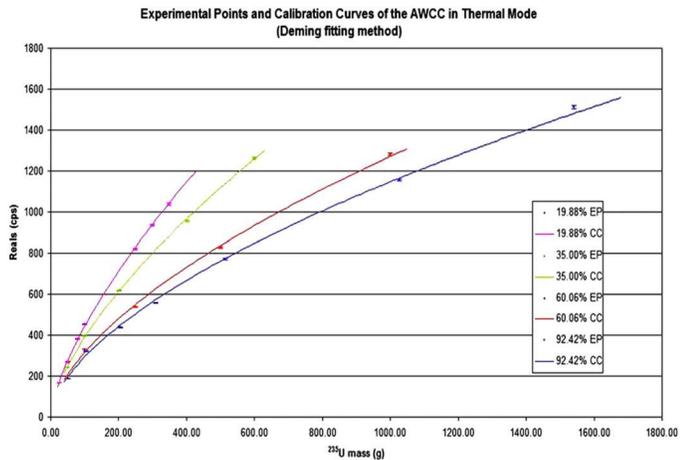

**Fig. 5.** Calibration curves from the active neutron well counter operated in thermal mode. Large discrepancies are visible between the calibration lines of various U-235 enrichments. This is because the coincidence count rate of the standard sample is highly sensitive to neutron multiplication. The high neutron multiplication sensitivity arises from the scattering of moderated neutrons in the polyethylene sheets back into the sample. The sample is not just interrogated by the AmLi neutron source but also by fission and source neutrons that scatter back into the source from the surrounding polyethylene sheets. Credit: F. Ferrari et al. [6]

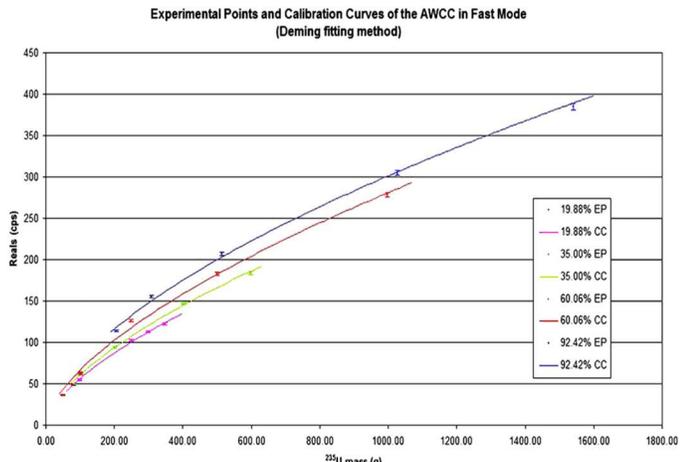

**Fig. 6.** Calibration curves from the active neutron well counter operated in fast mode. Cadmium sheets prevent the backscattering of neutrons thermalized for He3 detection. This reduces sensitivity of the coincidence count rate on the neutron multiplicity. The strong penetration of fast neutrons into the sample further reduce the dependency of the coincidence count rate on sample density and mass when compared to the thermal case. The neutron count rate is significantly lower than the thermal case rendering small samples difficult to measure. Credit: F. Ferrari et al. [6]

The performance of the active well counter is determined as a function of measurement time. Well known samples are utilized for the performance determination. The performance parameter is the relative error of the mass of the assay as a function of the declared 235 U mass. This is conducted for both the thermal and the fast mode. The results are shown in Figure 7 for the thermal mode for 92.42% enriched U-235 samples of different masses. Figure 8 shows the results for the fast mode for the identical assay enrichment [6].

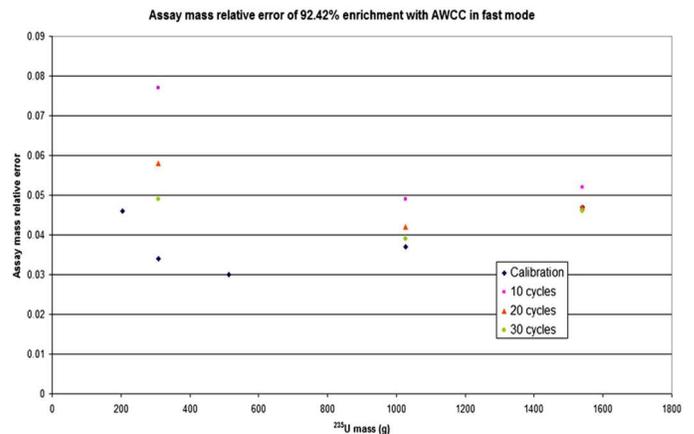

**Fig. 7.** The relative error of the mass of the assay as a function of the declared 92.42% enriched U-235 assay mass. Thermal mode is used. Credit: F. Ferrari et al. [6]

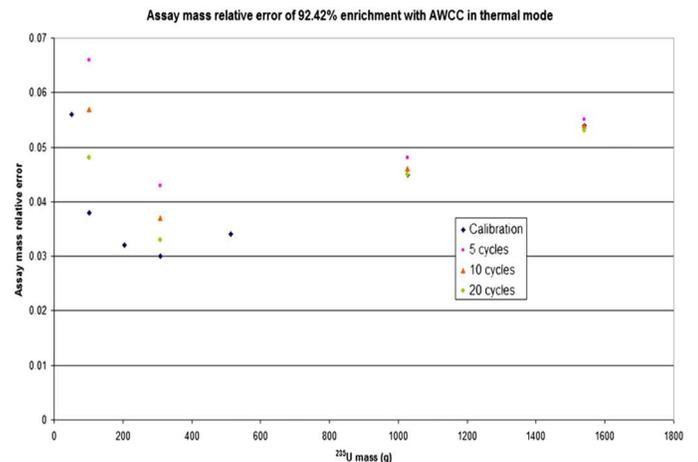

**Fig. 8.** The relative error of the mass of the assay as a function of the declared 92.42% enriched U-235 assay mass. Fast mode is used. Credit: F. Ferrari et al. [6]

Figure 7 and Figure 8 imply that the relative errors in mass of short and long measurement times begin to overlap for large 235 U masses. Further, at low assay masses, longer measurement times decrease the relative error in mass by nearly a factor of two. Lastly, for a given mass, significantly longer measurement times may not always lead to significantly lower relative errors in assay masses [6].

Based on this information, the researchers determined optimized measurement times. These optimized measurement times are shown in Table I and Table II [6].

TABLE I
OPTIMIZED MEASUREMENT TIMES FOR AN ACTIVE NEUTRON WELL COUNTER IN THERMAL MODE.
CREDIT: F. FERRARI ET AL. [6]

| U-235 Mass Rage | Recommended Measurement Time (s) |
|---|---|
| M < 200g | 2 000 |
| 200 < M < 500g | 1 000 |
| M > 500g | 5000 |

TABLE II
OPTIMIZED MEASUREMENT TIMES FOR AN ACTIVE NEUTRON WELL COUNTER IN FAST MODE.
CREDIT: F. FERRARI ET AL. [6]

| U-235 Mass Rage | Recommended Measurement Time |
|---|---|
| M < 300g | 3 000 |
| 300 < M < 1 000g | 2 000 |
| M > 1 000g | 1 000 |

V. CONCLUSION

Neutron well counters using the coincidence detection technique are widely employed by nuclear safeguards for verification of mass of declared fissile materials. Singlet counting may be possible for well-known assay samples. Calibration curves are required to determine masses in coincidence counting. Multiplicity counting enables assay mass determination without calibration curves, however calibration is still conducted for verification. Passive neutron well counters are well suited for determination of 240 Pu masses due to the large spontaneous fission cross sections of even mass numbered Plutonium isotopes. Active neutron well counters are well suited for determination of fissile material masses. Typical assay mass relative errors in HEU are 4% for assay samples above 1kg and 8% for assay samples of 50g. Thermal neutron interrogation sources are found to decrease relative errors in small samples. Measurement of large assay samples benefits from fast neutron interrogation sources due to lower self-multiplication and higher sample penetration.